\pdfoutput=1
\documentclass[12pt,a4paper]{article}
\usepackage{ifthen} % for conditional statements
\newboolean{pdflatex}
\setboolean{pdflatex}{true} % False for eps figures 

\newboolean{articletitles}
\setboolean{articletitles}{true} % False removes titles in references

\newboolean{uprightparticles}
\setboolean{uprightparticles}{false} %True for upright particle symbols

\usepackage{longtable} % only for template; not usually to be used in PAPERs
\usepackage{microtype}
\usepackage{lineno}  % for line numbering during review
\usepackage{xspace} % To avoid problems with missing or double spaces after
\usepackage{caption} %these three command get the figure and table captions automatically small

\usepackage{graphicx}  % to include figures (can also use other packages)
\usepackage{color}
\usepackage{colortbl}
\graphicspath{{./figs/}} % Make Latex search fig subdir for figures

\usepackage{amsmath} % Adds a large collection of math symbols
\usepackage{amssymb}
\usepackage{amsfonts}
\usepackage{upgreek} % Adds in support for greek letters in roman typeset

\newcommand*\patchAmsMathEnvironmentForLineno[1]{%
\expandafter\let\csname old#1\expandafter\endcsname\csname #1\endcsname
\expandafter\let\csname oldend#1\expandafter\endcsname\csname
end#1\endcsname
 \renewenvironment{#1}%
   {\linenomath\csname old#1\endcsname}%
   {\csname oldend#1\endcsname\endlinenomath}%
}
\newcommand*\patchBothAmsMathEnvironmentsForLineno[1]{%
  \patchAmsMathEnvironmentForLineno{#1}%
  \patchAmsMathEnvironmentForLineno{#1*}%
}
\AtBeginDocument{%
\patchBothAmsMathEnvironmentsForLineno{equation}%
\patchBothAmsMathEnvironmentsForLineno{align}%
\patchBothAmsMathEnvironmentsForLineno{flalign}%
\patchBothAmsMathEnvironmentsForLineno{alignat}%
\patchBothAmsMathEnvironmentsForLineno{gather}%
\patchBothAmsMathEnvironmentsForLineno{multline}%
\patchBothAmsMathEnvironmentsForLineno{eqnarray}%
}

\usepackage{hyperref}    % Hyperlinks in references
\usepackage[all]{hypcap} % Internal hyperlinks to floats.

\def\lhcb {\mbox{LHCb}\xspace}
\def\ux85 {\mbox{UX85}\xspace}

\ifthenelse{\boolean{uprightparticles}}%
{

 \def\Ppsi        {\ensuremath{\uppsi}\xspace}

 \def\PDelta      {\ensuremath{\Delta}\xspace}                 
 \def\PXi      {\ensuremath{\Xi}\xspace}                 
 \def\PLambda      {\ensuremath{\Lambda}\xspace}                 
 \def\PSigma      {\ensuremath{\Sigma}\xspace}                 
 \def\POmega      {\ensuremath{\Omega}\xspace}                 
 \def\PUpsilon      {\ensuremath{\Upsilon}\xspace}                 
 
 \def\PB      {\ensuremath{\mathrm{B}}\xspace}                 
                  
 \def\PD      {\ensuremath{\mathrm{D}}\xspace}

 \def\PJ      {\ensuremath{\mathrm{J}}\xspace}                 
 \def\PK      {\ensuremath{\mathrm{K}}\xspace}

 \def\Pi      {\ensuremath{\mathrm{i}}\xspace}

}
{

 \def\Ppsi        {\ensuremath{\psi}\xspace}                 
                  
 \mathchardef\PDelta="7101
 \mathchardef\PXi="7104
 \mathchardef\PLambda="7103
 \mathchardef\PSigma="7106
 \mathchardef\POmega="710A
 \mathchardef\PUpsilon="7107
                  
 \def\PB      {\ensuremath{B}\xspace}                 
                  
 \def\PD      {\ensuremath{D}\xspace}

 \def\PJ      {\ensuremath{J}\xspace}                 
 \def\PK      {\ensuremath{K}\xspace}

 \def\Pi      {\ensuremath{i}\xspace}

}

\def\pipi  {\ensuremath{\pion^+\pion^-}\xspace}

\def\kaon  {\ensuremath{\PK}\xspace}
  \def\Kbar  {\kern 0.2em\overline{\kern -0.2em \PK}{}\xspace}

\def\Kz    {\ensuremath{\kaon^0}\xspace}
\def\Kzb   {\ensuremath{\Kbar^0}\xspace}
\def\KzKzb {\ensuremath{\Kz \kern -0.16em \Kzb}\xspace}
\def\Kp    {\ensuremath{\kaon^+}\xspace}
\def\Km    {\ensuremath{\kaon^-}\xspace}

\def\KpKm  {\ensuremath{\Kp \kern -0.16em \Km}\xspace}

\def\Kstar   {\ensuremath{\kaon^*}\xspace}

  \def\Dbar    {\kern 0.2em\overline{\kern -0.2em \PD}{}\xspace}
\def\D       {\ensuremath{\PD}\xspace}

\def\Dz      {\ensuremath{\D^0}\xspace}
\def\Dzb     {\ensuremath{\Dbar^0}\xspace}
\def\DzDzb   {\ensuremath{\Dz {\kern -0.16em \Dzb}}\xspace}
\def\Dp      {\ensuremath{\D^+}\xspace}
\def\Dm      {\ensuremath{\D^-}\xspace}

\def\DpDm    {\ensuremath{\Dp {\kern -0.16em \Dm}}\xspace}

\def\Bbar    {\ensuremath{\kern 0.18em\overline{\kern -0.18em \PB}{}}\xspace}

\def\jpsi     {\ensuremath{{\PJ\mskip -3mu/\mskip -2mu\Ppsi\mskip 2mu}}\xspace}

  \def\Y#1S{\ensuremath{\PUpsilon{(#1S)}}\xspace}% no space before {...}!

\def\L {\ensuremath{\PLambda}\xspace}
\def\Lbar {\ensuremath{\kern 0.1em\overline{\kern -0.1em\PLambda}}\xspace}

\def\BR         {\BF}
\def\to                 {\ensuremath{\rightarrow}\xspace}

\def\AT#1     {\ensuremath{A_{\mathrm{T}}^{#1}}\xspace}           % 2

\def\C#1      {\ensuremath{\mathcal{C}_{#1}}\xspace}                       % 9
\def\Cp#1     {\ensuremath{\mathcal{C}_{#1}^{'}}\xspace}                    % 7
\def\Ceff#1   {\ensuremath{\mathcal{C}_{#1}^{\mathrm{(eff)}}}\xspace}        % 9  
\def\Cpeff#1  {\ensuremath{\mathcal{C}_{#1}^{'\mathrm{(eff)}}}\xspace}       % 7
\def\Ope#1    {\ensuremath{\mathcal{O}_{#1}}\xspace}                       % 2
\def\Opep#1   {\ensuremath{\mathcal{O}_{#1}^{'}}\xspace}                    % 7

\newcommand{\tev}{\ensuremath{\mathrm{\,Te\kern -0.1em V}}\xspace}
\newcommand{\gev}{\ensuremath{\mathrm{\,Ge\kern -0.1em V}}\xspace}
\newcommand{\mev}{\ensuremath{\mathrm{\,Me\kern -0.1em V}}\xspace}
\newcommand{\kev}{\ensuremath{\mathrm{\,ke\kern -0.1em V}}\xspace}
\newcommand{\ev}{\ensuremath{\mathrm{\,e\kern -0.1em V}}\xspace}
\newcommand{\gevc}{\ensuremath{{\mathrm{\,Ge\kern -0.1em V\!/}c}}\xspace}
\newcommand{\mevc}{\ensuremath{{\mathrm{\,Me\kern -0.1em V\!/}c}}\xspace}
\newcommand{\gevcc}{\ensuremath{{\mathrm{\,Ge\kern -0.1em V\!/}c^2}}\xspace}
\newcommand{\gevgevcccc}{\ensuremath{{\mathrm{\,Ge\kern -0.1em V^2\!/}c^4}}\xspace}
\newcommand{\mevcc}{\ensuremath{{\mathrm{\,Me\kern -0.1em V\!/}c^2}}\xspace}

\def\gsim{{~\raise.15em\hbox{$>$}\kern-.85em
          \lower.35em\hbox{$\sim$}~}\xspace}
\def\lsim{{~\raise.15em\hbox{$<$}\kern-.85em
          \lower.35em\hbox{$\sim$}~}\xspace}

\def\PDF {PDF\xspace}

\def\sPlot{\mbox{\em sPlot}}

\def\pt         {\mbox{$p_{\rm T}$}\xspace}

\def\tell1  {TELL1\xspace}
\def\ukl1   {UKL1\xspace}

\usepackage{cite} % Allows for ranges in citations
\usepackage{mciteplus}

\begin{document}

%%%%%%%%%%%%%%%%%%%%%%%%%
%%%%% Title     %%%%%%%%%
%%%%%%%%%%%%%%%%%%%%%%%%%
\renewcommand{\thefootnote}{\fnsymbol{footnote}}
\setcounter{footnote}{1}

% %%%%%%% CHOOSE TITLE PAGE--------
%\onecolumn
% \input{title-LHCb-ANA}
%\input{title-LHCb-CONF}
% $Id: title-LHCb-PAPER.tex 10646 2011-10-12 13:51:38Z uegede $
% ===============================================================================
% Purpose: LHCb-PAPER journal paper title page template
% Author: 
% Created on: 2010-09-25
% ===============================================================================

%%%%%%%%%%%%%%%%%%%%%%%%%
%%%%%  TITLE PAGE  %%%%%%
%%%%%%%%%%%%%%%%%%%%%%%%%
\begin{titlepage}
\pagenumbering{roman}

% Header ---------------------------------------------------
\vspace*{-1.5cm}
\centerline{\large EUROPEAN ORGANIZATION FOR NUCLEAR RESEARCH (CERN)}
\vspace*{1.5cm}
\hspace*{-0.5cm}
\begin{tabular*}{\linewidth}{lc@{\extracolsep{\fill}}r}
\ifthenelse{\boolean{pdflatex}}% Logo format choice
{\vspace*{-2.7cm}\mbox{\!\!\!\includegraphics[width=.14\textwidth]{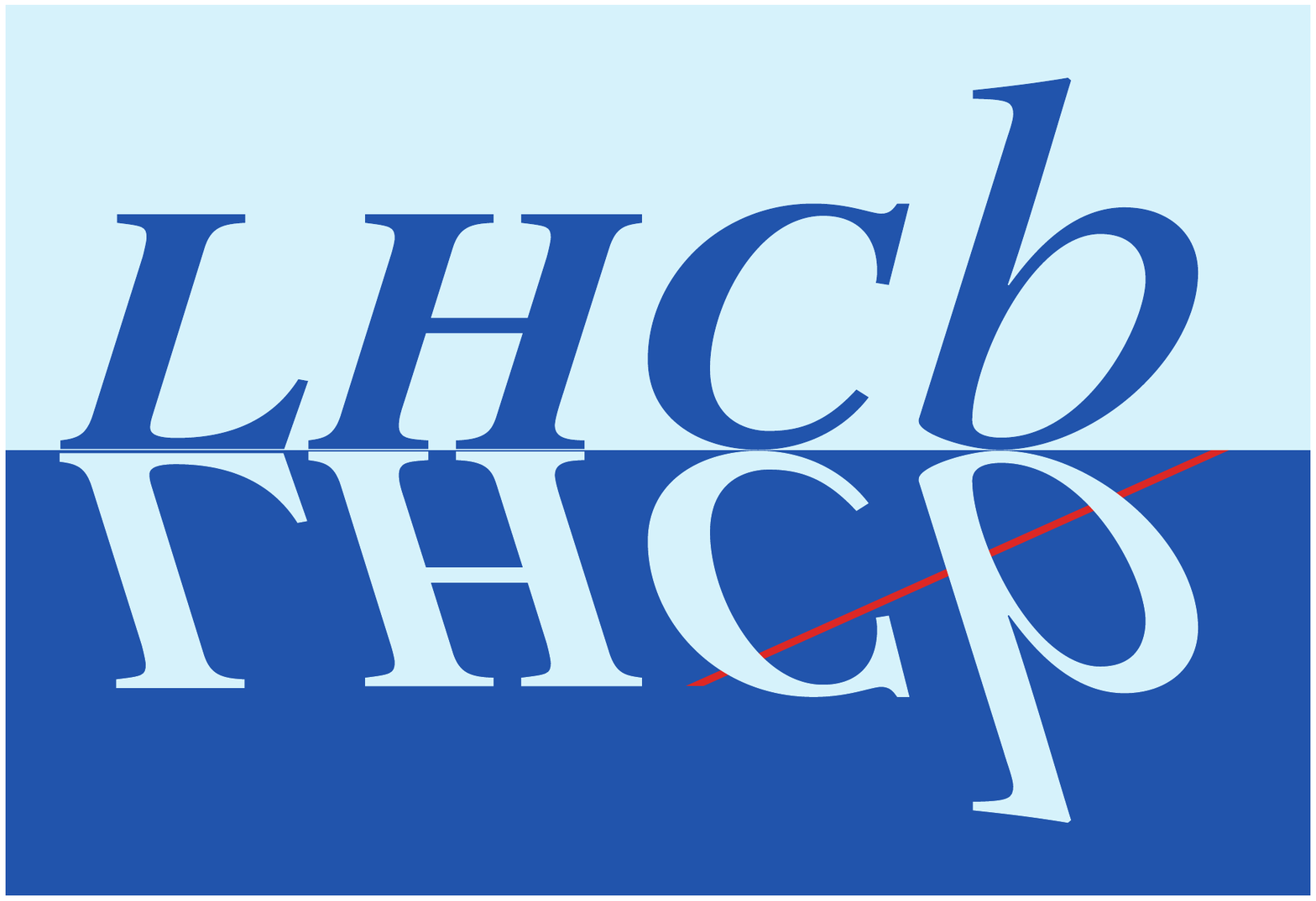}} & &}%
{\vspace*{-1.2cm}\mbox{\!\!\!\includegraphics[width=.12\textwidth]{lhcb-logo.eps}} & &}%
\\
 & & CERN-PH-EP-2015-098 \\  % ID 
 & & LHCb-PAPER-2015-015 \\  % ID 
 & & 30 June 2015 \\ % \today \\ %1 May 2014 \\  % Date - Can also hardwire e.g.: 23 March 2010
 & & \\
% not in paper \hline
\end{tabular*}

\vspace*{4.0cm}

% Title --------------------------------------------------
{\bf\boldmath\huge
\begin{center}
Quantum numbers of the $X(3872)$ state and orbital angular momentum in its $\rho^0\jpsi$ decay 
\end{center}
}

\vspace*{1.0cm}

% Authors -------------------------------------------------
\begin{center}
The LHCb collaboration\footnote{Authors are listed at the end of this paper.}
\end{center}

\vspace{\fill}

% Abstract -----------------------------------------------
\begin{abstract}
  \noindent
Angular correlations in $B^+\to X(3872) K^+$ decays, 
with $X(3872)\to \rho^0 J/\psi$, $\rho^0\to\pi^+\pi^-$  and $J/\psi \to\mu^+\mu^-$,
are used to measure orbital angular momentum contributions 
and to determine the $J^{PC}$ value of the $X(3872)$ meson.
The data correspond to an integrated luminosity of 
3.0 fb$^{-1}$ of proton-proton collisions collected with the LHCb detector. 
This determination, for the first time performed without assuming
a value for the orbital angular momentum, confirms the quantum
numbers to be $J^{PC}=1^{++}$.
The $X(3872)$ is found to decay predominantly through S wave and 
an upper limit of $4\%$ at $95\%$ C.L.\ is set on the D--wave contribution.
\end{abstract}

\vspace*{2.0cm}

\begin{center}
  Published in Phys. Rev. D92, 011102 (2015).
\end{center}

\vspace{\fill}

{\footnotesize 
\centerline{\copyright~CERN on behalf of the \lhcb collaboration, license \href{http://creativecommons.org/licenses/by/4.0/
}{CC-BY-4.0}.}}
\vspace*{2mm}

\end{titlepage}

%%%%%%%%%%%%%%%%%%%%%%%%%%%%%%%%
%%%%%  EOD OF TITLE PAGE  %%%%%%
%%%%%%%%%%%%%%%%%%%%%%%%%%%%%%%%

%  empty page follows the title page ----
\newpage
\setcounter{page}{2}
\mbox{~}
\newpage

% Author List ----------------------------
%  You need to get a new author list!
%\input{LHCb_authorlist.tex}

\cleardoublepage

%\twocolumn
% %%%%%%%%%%%%% ---------

\renewcommand{\thefootnote}{\arabic{footnote}}
\setcounter{footnote}{0}

%%%%%%%%%%%%%%%%%%%%%%%%%%%%%%%%
%%%%%  Table of Content   %%%%%%
%%%%%%%%%%%%%%%%%%%%%%%%%%%%%%%%
%%%% Uncomment next 2 lines if desired
%\tableofcontents
%\cleardoublepage

%%%%%%%%%%%%%%%%%%%%%%%%%
%%%%% Main text %%%%%%%%%
%%%%%%%%%%%%%%%%%%%%%%%%%

\pagestyle{plain} % restore page numbers for the main text
\setcounter{page}{1}
\pagenumbering{arabic}

%% Uncomment during review phase. 
%% Comment before a final submission.
%\linenumbers

% You can include short sections directly in the main tex file.
% However, for larger papers it is desirable to split the text into
% several semiautonomous files, which can be revised independently.
% This is especially useful when developing a document in
% collaboration with several people, since then different parts can be
% edited independently.  This type of file organization is shown here.
% 

\newboolean{prl}
\setboolean{prl}{false} % False for eps figures 

\newlength{\figsize}
\setlength{\figsize}{0.9\hsize}
% --------------------
\def\bpsi2skp{\bar{B}^0\to\psi(2S)K^-\pi^+}
\def\bjpsikp{\bar{B}^0\to\jpsi K^-\pi^+}
\def\bjpsiks{\bar{B}^0\to\jpsi K^*}
\def\bpsiks{\bar{B}^0\to\psi K^*}
\def\bchic1kp{\bar{B}^0\to\chi_{c1} K^-\pi^+}
\def\bcjppp{B_c^+\to\jpsi\pi^+\pi^-\pi^+}
\def\bjkkk{B^+\to\jpsi K^+K^-K^+}
\def\bjkpp{B^+\to\jpsi K^+\pi^+\pi^-}
\def\Kstar{K^*}
\def\BR{{\cal B}}
\def\L{{\cal L}}
\def\DLL{DLL}
\def\PDF{{\cal P}}
\def\M{{\cal M}}
\def\NDOF{\hbox{\rm ndf}}
\def\cospsi{\cos\theta_{\psi}}
\def\cospsiz{\cos\theta_{\psi}^Z}
\def\cosks{\cos\theta_{K^*}}
\def\cosz{\cos\theta_{Z}}
\def\mkp{m_{K\pi}}
\def\mkpi{\mkp}
\def\mpsi2sp{m_{\psi(2S)\pi}}
\def\mpsip{m_{\psi\pi}}
\def\mpsipi{\mpsip}
\def\Sum{\sum}
\def\Int{\int}
\def\Frac{\frac}
\def\xff{X(3872)}
\def\bpsi2sk{B^+\to\psi(2S) K^+}
\def\bxk{B^+\to\xff K^+}
\def\bjks{B^+\to\jpsi K^{*+}}
\def\kone{K_1(1270)^+}
\def\bjkone{B^+\to\jpsi\kone}
\def\xppj{\xff\to\pi^+\pi^-\jpsi}
\def\xrj{\xff\to\rho^0\jpsi}
\def\rpp{\rho^0\to\pi^+\pi^-}
\def\pipi{\pi\pi}
\def\jll{\jpsi\to \ell^+\ell^-}
\def\jmm{\jpsi\to \mu^+\mu^-}
\def\jpc{J^{PC}}
\def\DM{\Delta M}
\def\mjjp{\DM} %M(\pi^+\pi^-\jpsi)-M(\jpsi)}
\def\mppj{\DM} %M(\pi^+\pi^-\jpsi)-M(\jpsi)}
\def\mjppj{\DM} %M(\pi^+\pi^-\jpsi)-M(\jpsi)}
\def\jppk{\jpsi\pi^+\pi^-K^+}
\def\jkpp{\jpsi\pi^+\pi^-K^+}
\def\pp{\pi\pi}
\def\mpp{M(\pi^+\pi^-)}
\def\Q{M(\jpsi\pi^+\pi^-)-M(\jpsi)-\mpp}
\def\L{{\cal L}}
\def\sWeights{{\mbox{\em sWeights}}}
\def\Lmin{L_{\rm min}}
% -------------------------

\noindent
The $\xff$ state 
was discovered in 
$B^{+,0}\to\xff K^{+,0}$, 
$\xff\to\pi^+\pi^-\jpsi$, 
$\jpsi\to\ell^+\ell^-$  decays
by the Belle experiment 
\cite{Choi:2003ue} and
subsequently confirmed by other
experiments \cite{CDFPhysRevLett.93.072001,D0Abazov:2004kp,BaBarPhysRevD.71.071103}.\footnote{ 
The inclusion of charge-conjugate states is implied in this article.}
Its production was also studied at the LHC \cite{Aaij:2011sn,Chatrchyan:2013cld}.
However, the nature of this state remains unclear. 
The $\xff$ state is narrow, 
has a mass very close to 
the $D^0\Dbar^{*0}$ 
threshold 
and decays to $\rho^0\jpsi$ and $\omega\jpsi$ final states
with comparable branching fractions \cite{PDG2014}, 
thus violating isospin symmetry.
This suggests that the $\xff$ particle may not be a simple $c\bar{c}$ state, 
and exotic states such as 
$D^0\Dbar^{*0}$ molecules \cite{Tornqvist:2004qy}, 
tetraquarks \cite{Maiani:2004vq} or mixtures of states \cite{Hanhart:2011jz}
have been proposed to explain its composition.
The $X(3872)$ quantum numbers, 
such as total angular momentum $J$, parity $P$, 
and charge-conjugation $C$, 
impose constraints on the theoretical models of this state.  
The orbital angular momentum $L$ in the $\xff$ decay may also provide
information on its internal structure. 

Observations of the $\xff\to\gamma\jpsi$ and $\xff\to\gamma\psi(2S)$  decays 
\cite{Aubert:2006aj,Bhardwaj:2011dj,LHCb-PAPER-2014-008} imply positive $C$, which
requires the total angular momentum
of the dipion system ($J_{\pi\pi}$) in $\xff\to\pi^+\pi^-\jpsi$ decays to be odd.
Dipion mass, $\mpp$, is limited by the available phase space 
to be less than $775$ MeV, and so 
$J_{\pi\pi}\geq 3$ can be ruled out 
since there are no known or predicted mesons 
with such high spins at such low masses.\footnote{ 
We use mass and momentum units in which $c=1$.}  
In fact, the distribution of $\mpp$ is consistent 
with $\xff\to\rho^0\jpsi$ decays \cite{Choi:2011fc,Abulencia:2005zc,Chatrchyan:2013cld},
%conforming with $J_{\pi\pi}=1$, the only plausible value.
in line with $J_{\pi\pi}=1$, the only plausible value.

The choices for $J^{PC}$ were narrowed down 
to two possibilities, $1^{++}$ or $2^{-+}$, by the CDF collaboration,
via an analysis of 
the angular correlations in inclusively reconstructed $\xff\to\pi^+\pi^-\jpsi$ and $\jpsi\to\mu^+\mu^-$ 
decays, dominated by prompt production in $p\bar p$ collisions \cite{Abulencia:2006ma}.
Using 1.0 fb$^{-1}$ of $p p$ collision data collected by LHCb, 
$J^{PC}=2^{-+}$ was ruled out in favor of the $1^{++}$
assignment, using the angular correlations in the same decay chain, with the $\xff$
state produced in $B^+\to\xff K^+$ decays \cite{LHCb-PAPER-2013-001}.
Both angular analyses assumed that the lowest orbital
angular momentum between the $\xff$ decay products ($\Lmin$) dominated the matrix element.
Significant contributions from $\Lmin+2$ amplitudes could invalidate the $1^{++}$ assignment.  
Since the phase-space limit on $\mpp$ is close to the $\rho^0$ pole 
($775.3\pm0.3$ MeV \cite{PDG2014}),
the energy release in the $\xff$ decay, $Q\equiv\Q$, is a small fraction of the
$\xff$ mass, making the orbital angular momentum barrier effective.\footnote{
Dimuon candidates are constrained to the known $\jpsi$ mass \cite{PDG2014}.}
However, an exotic component in $\xff$ could  
induce contributions from higher orbital angular 
momentum for models in which the size of the $\xff$ state is 
substantially larger than the compact sizes of the charmonium states.    
Therefore, it is important to probe the $\xff$ spin-parity without any assumptions about $L$.
A determination of the magnitude of 
contributions from $\Lmin+2$ amplitudes 
for the correct $J^{PC}$ is also of interest, since a substantial value
would suggest an anomalously large size of the $\xff$ state.
In this article, we extend our previous analysis \cite{LHCb-PAPER-2013-001}
of five-dimensional angular correlations in 
$B^+\to\xff K^+$, $\xff\to\rho^0\jpsi$, $\rho^0\to\pi^+\pi^-$, $\jmm$
decays to accomplish these goals. The integrated luminosity of the data sample has been tripled by 
adding $8$~TeV $pp$ collision data 
collected in 2012. 

\begin{figure}[tbp]
  \begin{center}
  \ifthenelse{\boolean{pdflatex}}{ 
        \includegraphics*[width=\figsize]{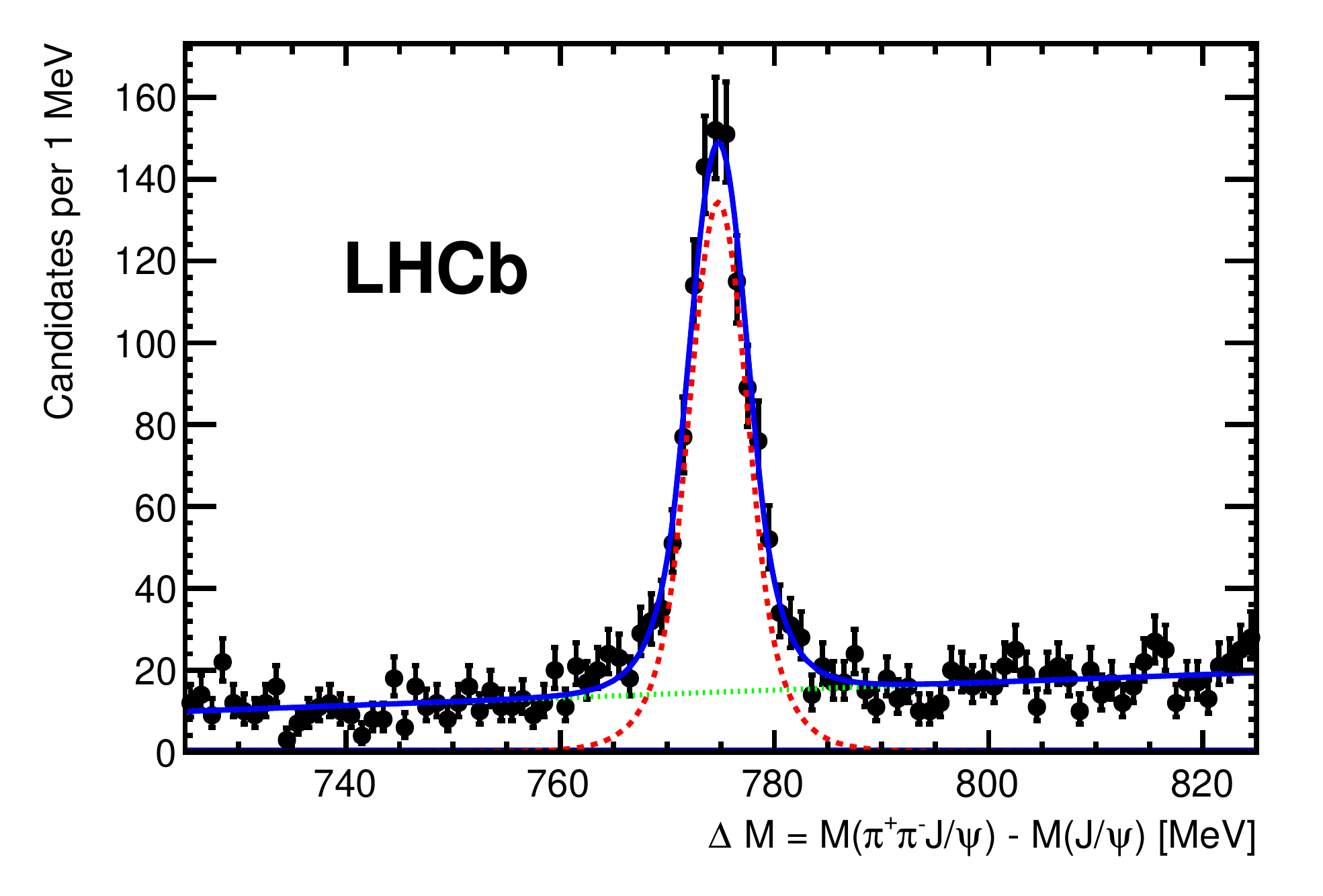}
   }{
        \includegraphics*[width=\figsize]{jppj_fit_x3872.eps}
   } 
  \end{center}
  \vskip-1.1cm\caption{\small 
    Distribution of $\mppj$ for $\bjkpp$ candidates.
    The fit of the $\xff$ signal is displayed.
    The solid (blue), dashed (red), and dotted (green) lines represent 
    the total fit, signal component, and background component, 
    respectively.
  \label{fig:jppj}
  }
\end{figure}

The \lhcb detector is a single-arm forward spectrometer covering the 
pseudorapidity range \mbox{$2<\eta<5$}, described in detail in 
Ref.~\cite{Alves:2008zz,LHCb-DP-2014-002}.
The $\xff$ candidate selection, which is based on reconstructing 
$B^+\to(\jpsi\to\mu^+\mu^-)\pi^+\pi^-K^+$ 
candidates using particle identification information, 
transverse momentum ($\pt$) thresholds and 
requiring separation of tracks and the $B^+$ vertex  
from the primary $pp$ interaction vertex, 
is improved relative to that of Ref.~\cite{LHCb-PAPER-2013-001}.
The signal efficiency is increased by lowering 
requirements on $\pt$ 
for muons from 0.90 to 0.55 GeV and for hadrons from 0.25 to 0.20 GeV. 
The background is further suppressed without 
significant loss of signal by requiring $Q<250$ MeV.
The $\xff$ mass resolution ($\sigma_{\Delta M}$)
is improved from about $5.5$ MeV to $2.8$ MeV
by constraining
the $B^+$ candidate to its known mass and requiring its momentum to
point to a $pp$ collision vertex in the kinematic fit of its decay.
The distribution of $\Delta M\equiv M(\pi^+\pi^-\jpsi)-M(\jpsi)$ 
is shown in Fig.~\ref{fig:jppj}.
A Crystal Ball function \cite{Skwarnicki:1986xj} with symmetric tails is used 
to model the signal shape, while
the background is assumed to be linear.
An unbinned maximum likelihood fit yields  $1011\pm38$ $\bxk$ decays
and $1468\pm44$ background entries in the 
$725<\Delta M<825$ MeV range used in the angular analysis. 
The signal purity is $80\%$ within $2.5\,\sigma_{\Delta M}$ 
from the signal peak.   
From studying the $K^+\pi^+\pi^-$ mass distribution,
the dominant source of background is found to be 
$B^+\to\jpsi K_1(1270)^+$, $K_1(1270)^+\to K^+\pi^+\pi^-$ decays.

Angular correlations in the $B^+$ decay chain  
are analyzed using an unbinned maximum-likelihood fit
to determine the $\xff$ quantum numbers and orbital angular momentum in its decay.
The probability density function ($\PDF$) for each $\jpc$ hypothesis, $J_X$, is defined
in the five-dimensional angular space $\Omega\equiv$
$(\cos\theta_X,\cos\theta_{\rho},\Delta\phi_{X,\rho},\cos\theta_{\jpsi},\Delta\phi_{X,\jpsi})$,
where $\theta_X$, $\theta_{\rho}$ and $\theta_{\jpsi}$ 
are the helicity angles \cite{Jacob:1959at,Richman:1984gh,PhysRevD.57.431}
in $\xff$, $\rho^0$ and $\jpsi$ decays, respectively,
and $\Delta\phi_{X,\rho}$, $\Delta\phi_{X,\jpsi}$ are the angles 
between the decay planes of the $\xff$ particle and of its decay products.  
The quantity $\PDF$ is 
the normalized product of
the expected decay matrix element ($\M$) squared
and of the reconstruction efficiency ($\epsilon$), 
$\PDF(\Omega|J_X)=|\M(\Omega|J_X)|^2\,\epsilon(\Omega)/I(J_X)$, where
$I(J_X)=\int|\M(\Omega|J_X)|^2\,\epsilon(\Omega){\it d}\Omega$.
The efficiency is averaged over the $\pi^+\pi^-$ mass using  
a simulation \cite{Sjostrand:2006za,LHCb-PROC-2010-056,Lange:2001uf,Allison:2006ve,*Agostinelli:2002hh,LHCb-PROC-2011-006}
of the $\xff\to\rho^0\jpsi$, $\rho^0\to\pi^+\pi^-$ decay. 
%\cite{Choi:2011fc,Abulencia:2005zc,Chatrchyan:2013cld}.
%The observed $\mpp$ distribution is in agreement with this simulation.
The lineshape of the $\rho^0$ resonance can change slightly depending on
the $\xff$ spin hypothesis.
The effect on $\epsilon(\Omega)$ is very small and is neglected. 
The angular correlations are obtained using the helicity formalism \cite{Abulencia:2006ma},
\ifthenelse{\boolean{prl}}{ 
\begin{eqnarray}
 |\, \M(\Omega|J_X)\,|^2 & = & \sum_{\Delta\lambda_{\mu}=-1,+1}  \left| \phantom{A_{\lambda_\psi}}\hskip-0.4cm \right.
 \sum_{\lambda_{\jpsi},\lambda_{\rho}=-1,0,+1}  \notag\\
A_{\lambda_{\jpsi},\lambda_{\rho}} &  %\!\!\!\!\!\!\!\!\!
 %\times 
 & D^{J_X}_{0\,,\,\lambda_{\jpsi}-\lambda_{\rho}}(0,\theta_X,0)^* 
%\times 
\notag\\
& & D^{1}_{\lambda_{\rho}\,,\,0}(\Delta\phi_{X,\rho},\theta_{\rho},0)^* 
%\times 
\notag\\
& & 
D^{1}_{\lambda_{\jpsi}\,,\,\Delta\lambda_{\mu}}(\Delta\phi_{X,\jpsi},\theta_{\jpsi},0)^*
\left. \phantom{A_{\lambda_\psi}}\hskip-0.4cm \right|^2, \notag\\
\label{eq:m2}
\end{eqnarray}
}{
\begin{eqnarray}  |\, \M(\Omega|J_X)\,|^2 & = & \sum_{\Delta\lambda_{\mu}=-1,+1}  \left| \phantom{A_{\lambda_\psi}}\hskip-0.4cm \right.  
\sum_{\lambda_{\jpsi},\lambda_{\rho}=-1,0,+1} A_{\lambda_{\jpsi},\lambda_{\rho}} 
%\times D^{J_X}_{0\,,\,\lambda_{\jpsi}-\lambda_{\rho}}(\phi_X,\theta_X,-\phi_X) \times \notag\\
\quad %\times 
D^{J_X}_{0\,,\,\lambda_{\jpsi}-\lambda_{\rho}}(0,\theta_X,0)^* 
%\times 
\notag\\
 & & \hskip2.4cm 
%D^{1}_{\lambda_{\rho}\,,\,0}(\phi_{\rho},\theta_{\rho},-\phi_{\rho}) \times  D^{1}_{\lambda_{\jpsi}\,,\,\Delta\lambda_{\mu}}(\phi_{\jpsi},\theta_{\jpsi},-\phi_{\jpsi})  
 D^{1}_{\lambda_{\rho}\,,\,0}(\Delta\phi_{X,\rho},\theta_{\rho},0)^* 
\quad%\times 
D^{1}_{\lambda_{\jpsi}\,,\,\Delta\lambda_{\mu}}(\Delta\phi_{X,\jpsi},\theta_{\jpsi},0)^*
\left. \phantom{A_{\lambda_\psi}}\hskip-0.4cm \right|^2 , \notag\\  
\label{eq:m2}
\end{eqnarray}
}
\noindent
where $\lambda$ are particle helicities, $\Delta\lambda_\mu=\lambda_{\mu^+}-\lambda_{\mu^-}$,
and $D^J_{\lambda_1\,,\,\lambda_2}$ 
are Wigner functions \cite{Jacob:1959at,Richman:1984gh,PhysRevD.57.431}.
The helicity couplings, $A_{\lambda_{\jpsi},\lambda_{\rho}}$,
are expressed in terms of the $LS$ couplings, 
$B_{LS}$, with the help of Clebsch-Gordan coefficients, 
where $L$ is the orbital angular momentum between the $\rho^0$ 
and the $\jpsi$ mesons, and 
$S$ is the sum of their spins,
\ifthenelse{\boolean{prl}}{ 
\begin{eqnarray}
A_{\lambda_{\jpsi}\,,\,\lambda_{\rho}}=\sum_L \sum_S 
B_{LS} %\times 
& 
\left( 
\begin{array}{cc|c}
 J_{\jpsi} & J_{\rho} & S \\
 \lambda_{\jpsi} & -\lambda_{\rho} & \lambda_{\jpsi}-\lambda_{\rho} 
\end{array}
\right)
%\times 
\notag\\
&
\left( 
\begin{array}{cc|c}
 L  & S & J_X \\
 0 & \lambda_{\jpsi}-\lambda_{\rho} & \lambda_{\jpsi}-\lambda_{\rho}   
\end{array}
\right).
\notag\\
\label{eq:LS}
\end{eqnarray}
}{
\begin{eqnarray}
A_{\lambda_{\jpsi}\,,\,\lambda_{\rho}}=\sum_L \sum_S 
B_{LS} %\times %& 
\left( 
\begin{array}{cc|c}
 J_{\jpsi} & J_{\rho} & S \\
 \lambda_{\jpsi} & -\lambda_{\rho} & \lambda_{\jpsi}-\lambda_{\rho} 
\end{array}
\right)
%\times %\notag\\
%&
\left( 
\begin{array}{cc|c}
 L  & S & J_X \\
 0 & \lambda_{\jpsi}-\lambda_{\rho} & \lambda_{\jpsi}-\lambda_{\rho}   
\end{array}
\right).
\notag\\
\label{eq:LS}
\end{eqnarray}
}
Possible values of $L$ are constrained by parity conservation,
%\begin{equation}
$P_X=P_{\jpsi}\,P_{\rho}\,(-1)^L=(-1)^L$.
%\label{eq:L}
%\end{equation}
In the previous analyses \cite{Abulencia:2006ma,Choi:2011fc,LHCb-PAPER-2013-001}, 
only the minimal value of the angular
momentum, $\Lmin$, was allowed.
Thus, for the preferred $\jpc=1^{++}$
hypothesis, the D wave was neglected allowing only S--wave decays. 
In this work all $L$ values are allowed in Eq.~(\ref{eq:LS}).
The corresponding $B_{LS}$ amplitudes are listed in Table~\ref{tab:LScombo}.
Values of $J_X$ up to four are analyzed. Since the orbital angular momentum in 
the $B^+$ decay equals
$J_X$, high values are suppressed by the angular momentum barrier. In fact, the highest
observed spin of any resonance produced in $B$ decays is three \cite{LHCb-PAPER-2014-035,LHCb-PAPER-2014-036}.
Since $\PDF$ is insensitive to the overall normalization of the $B_{LS}$ couplings 
and to the phase of the matrix element, the $B_{LS}$ amplitude with the lowest 
$L$ and $S$ is set to the arbitrary reference value $(1,0)$. 
The set of other possible complex $B_{LS}$ amplitudes, which are free parameters in the 
fit, is denoted as $\alpha$.
\begin{table}[htbp]
\centering
\caption{Parity-allowed $LS$ couplings in the $\xff\to\rho^0\jpsi$ decay.
         For comparison, we also list a subset of these couplings corresponding to 
         the lowest $L$, used in the previous determinations \cite{Abulencia:2006ma,LHCb-PAPER-2013-001,Choi:2011fc} 
         of the $\xff$ quantum numbers.}
\label{tab:LScombo}
\begin{tabular}{lll}
\\
\hline
        &   \multicolumn{2}{c}{ $B_{LS}$ } \\
 $\jpc$ &   Any $L$ value  & Minimal $L$ value \\
\hline
 $0^{-+}$  & $B_{11}$                                                          & $B_{11}$ \\
 $0^{++}$ & $B_{00}, B_{22}$                                            & $B_{00}$ \\
 $1^{-+}$  & $B_{10}, B_{11}, B_{12}, B_{32}$                & $B_{10}, B_{11}, B_{12}$ \\
 $1^{++}$ & $B_{01}, B_{21}, B_{22}$                              & $B_{01}$ \\
 $2^{-+}$  & $B_{11}, B_{12}, B_{31}, B_{32}$                & $B_{11}, B_{12}$ \\
 $2^{++}$ & $B_{02}, B_{20}, B_{21}, B_{22}, B_{42}$  & $B_{02}$ \\
 $3^{-+}$  & $B_{12}, B_{30}, B_{31}, B_{32}, B_{52}$  & $B_{12}$ \\
 $3^{++}$ & $B_{21}, B_{22}, B_{41}, B_{42}$                & $B_{21}, B_{22}$ \\
 $4^{-+}$  & $B_{31}, B_{32}, B_{51}, B_{52}$                & $B_{31}, B_{32}$ \\
 $4^{++}$ & $B_{22}, B_{40}, B_{41}, B_{42}, B_{62}$  & $B_{22}$ \\
\hline
\end{tabular}
\end{table}

The function to be minimized is 
$-2\ln \L(J_X,\alpha) 
\equiv - s_w\,2\sum_{i=1}^{N_{\rm data}} w_i \ln \PDF(\Omega_i|J_X,\alpha)$,
where $\L(J_X,\alpha)$ is the unbinned likelihood, and
$N_{\rm data}$ is the number of selected candidates.
The background is subtracted 
using the \sPlot\ technique \cite{2005NIMPA.555..356P} 
by assigning a weight, % (\sWeight), 
$w_i$,
to each candidate 
 based on its $\Delta M$ value
(see Fig.~\ref{fig:jppj}).
No correlations between $\Delta M$ and $\Omega$ are observed.
Prompt production of $\xff$ in $pp$ collisions gives negligible contribution
to the selected sample.
Statistical fluctuations in the background subtraction are 
taken into account in the log-likelihood value via a constant
scaling factor,
$s_w = \sum_{i=1}^{N_{\rm data}} w_i/\sum_{i=1}^{N_{\rm data}} {w_i}^2$.
The efficiency $\epsilon(\Omega)$ is not
determined on an event-by-event basis, 
since it cancels in the likelihood ratio except for the normalization 
integrals.
A large sample of simulated events, with uniform angular distributions,
passed through a full simulation of the detection and the data 
selection process,
is used to carry out the integration, 
$I(J_X)\propto \sum_{i=1}^{N_{\rm MC}} |\M(\Omega_i|J_X)|^2$, 
where $N_{\rm MC}$ is the number of reconstructed simulated events.
The negative log-likelihood is minimized for each $J_X$ value
with respect to free $B_{LS}$ couplings, yielding their
estimated set of values $\hat{\alpha}$.
Hereinafter, $\L(J_X)\equiv\L(J_X,\hat{\alpha})$.  

\begin{figure}[tbp]
  \begin{center}
  \ifthenelse{\boolean{pdflatex}}{
    \includegraphics*[width=\figsize]{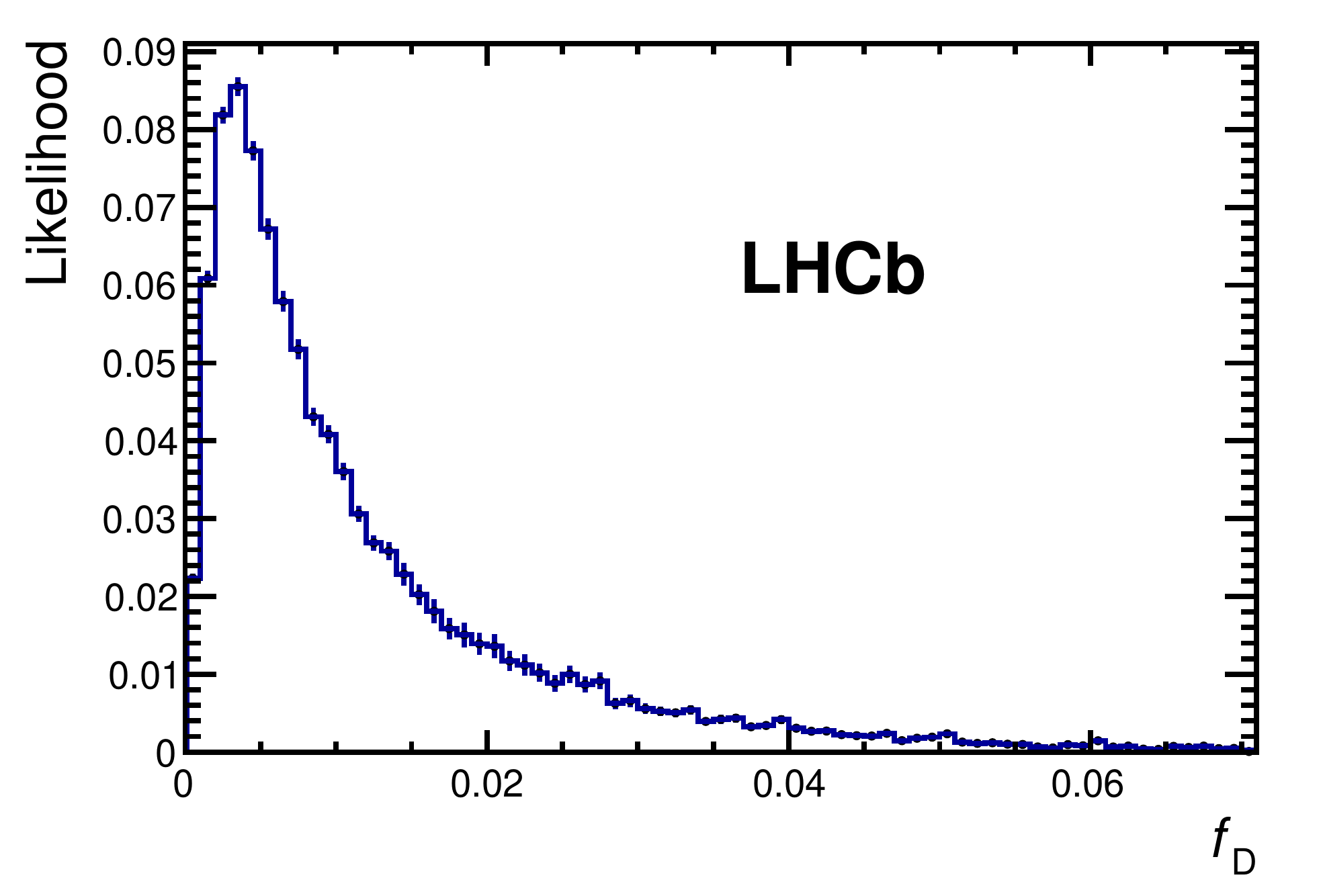}  
   }{
    \includegraphics*[width=\figsize]{dfracL.eps}
   } 
  \end{center}
  \vskip-0.1cm\caption{\small
  Likelihood-weighted distribution of the D--wave fraction. 
  The distribution is normalized to unity.
  \label{fig:dfracL}
  }
\end{figure}

The $1^{++}$ hypothesis gives the highest likelihood value.
From angular momentum and parity conservation, there are
two possible values of orbital angular momentum in the $\xff$ decay 
for this $J^{PC}$ value,
$L=0$ or $2$. 
For the S--wave decay, the total spin of the $\rho^0$ and $\jpsi$ mesons must be $S=1$;
thus $B_{01}$ is the only possible $LS$ amplitude. 
For the D--wave decay, two values are possible, $S=1$ or $2$,
corresponding to the amplitudes $B_{21}$ and $B_{22}$, respectively. 
The squared magnitudes of both of these D--wave amplitudes are consistent with zero, 
as demonstrated by the ratios
$|B_{21}|^2/|B_{01}|^2=0.002\pm0.004$
and 
$|B_{22}|^2/|B_{01}|^2=0.007\pm0.008$. 
Overall, the D--wave significance is only $0.8$ standard deviations
as obtained by applying Wilks theorem to 
the ratio of the likelihood values with the D--wave amplitudes floated in the fit 
and with them fixed to zero. 
The total D--wave fraction depends on the $B_{LS}$ amplitudes,
$f_{\rm D}\equiv \int \left| \M(\Omega)_{\rm D} \right|^2  d\Omega /
           \int \left| \M(\Omega)_{\rm S+D} \right|^2  d\Omega$, 
where $\M(\Omega)_{D}$ is the matrix element restricted to the  
$B_{21}$ and $B_{22}$ amplitudes only
and  $\M(\Omega)_{S+D}$ is the full matrix element. 
To set an upper limit on $f_{\rm D}$, we populate the four-dimensional 
space of complex $B_{21}$ and $B_{22}$ parameters with uniformly distributed 
points in a large region around the $B_{21}$ and $B_{22}$ fit values
($\pm14$ standard deviations in each parameter). 
For each point we determine the likelihood value from the data
and an $f_{\rm D}$ value via numerical integration of
the matrix element squared. 
The distribution of $f_{\rm D}$ values weighted 
by the likelihood values is shown in Fig.~\ref{fig:dfracL}. 
It peaks at $0.4\%$ with a non-Gaussian tail at higher values.
An upper limit of $f_{\rm D}<4\%$ at 95\%\ C.L.\ 
is determined using a Bayesian approach.

\begin{figure}[tbp]
  \begin{center}
  \ifthenelse{\boolean{pdflatex}}{
    \includegraphics*[width=\figsize]{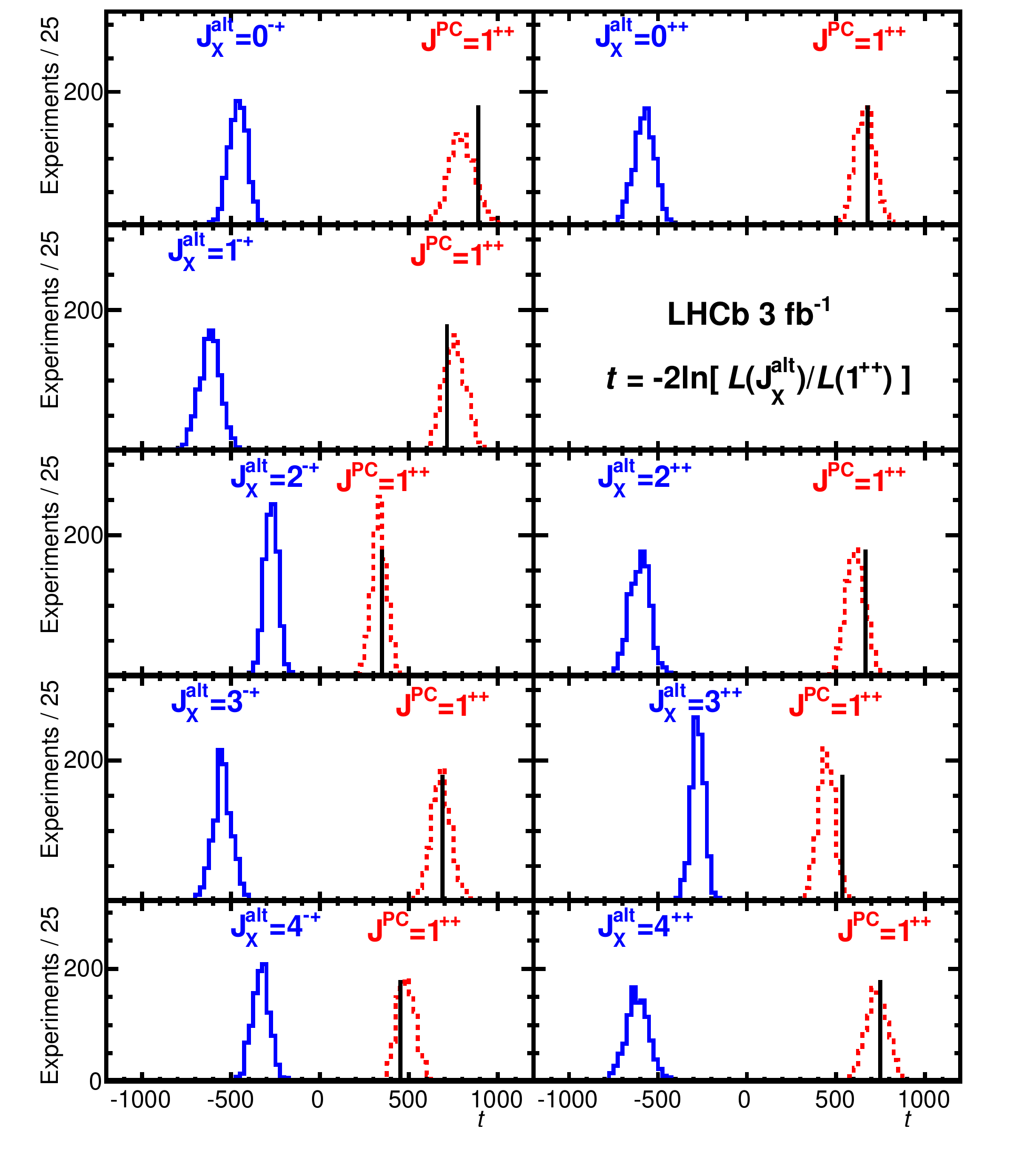}  
   }{
    \includegraphics*[width=\figsize]{spinana.eps}
   } 
  \end{center}
  \vskip-0.1cm\caption{\small
    Distributions of the test statistic $t \equiv -2\ln[\L(J_X^{\rm alt}))/\L(1^{++})]$,
    for the simulated experiments 
    under the $\jpc=J_X^{\rm alt}$ hypothesis
    (blue solid histograms)
    and under the $\jpc=1^{++}$ hypothesis (red dashed histograms).
    The values of the test statistics for the data, $t_{\rm data}$,
    are shown by the solid vertical lines. 
  \label{fig:spinana}
  }
\end{figure}

\begin{figure}[htbp]
  \begin{center}
  \ifthenelse{\boolean{pdflatex}}{
    \includegraphics*[width=\figsize]{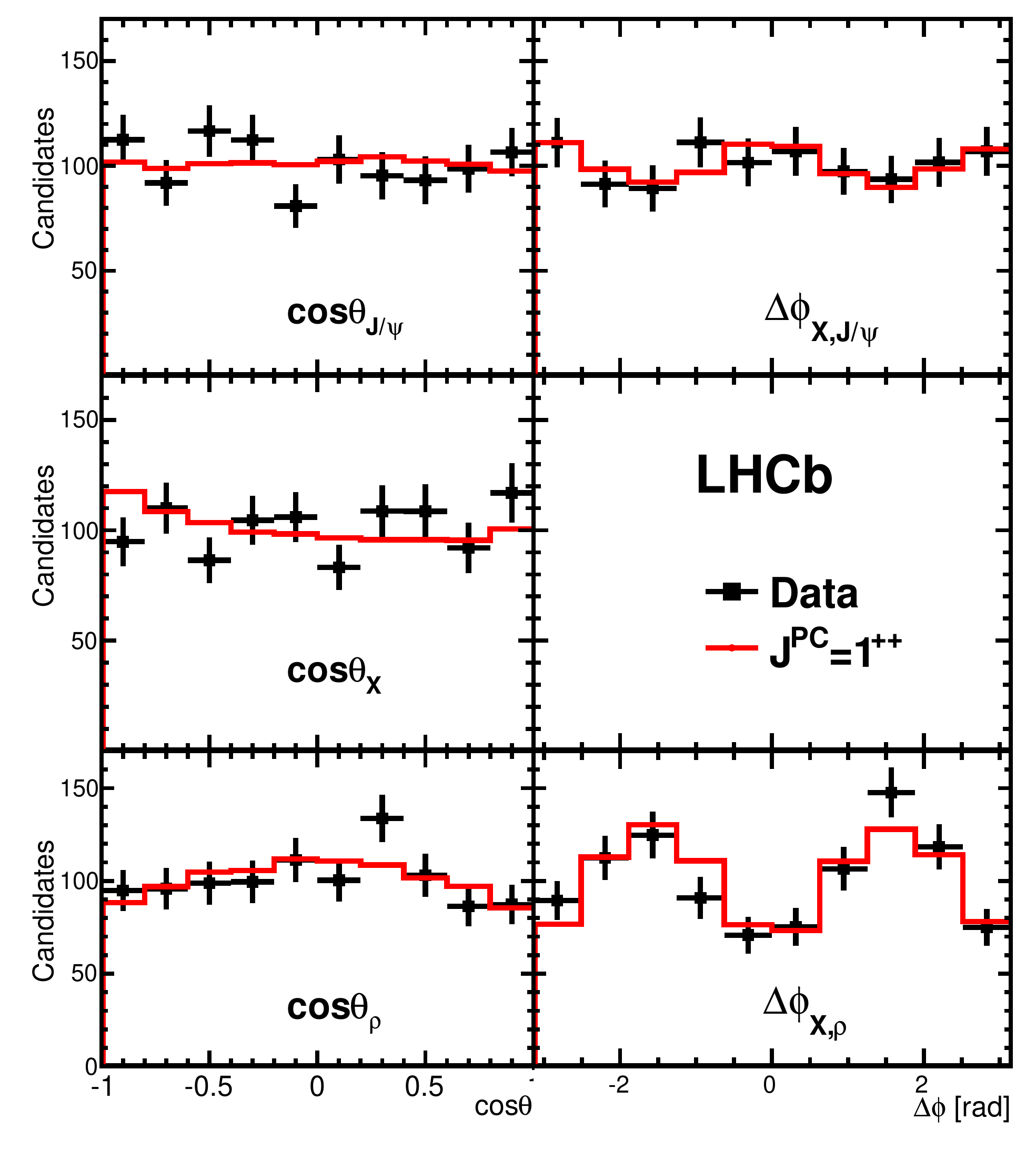}
   }{
    \includegraphics*[width=\figsize]{angles1P.eps}
   } 
  \end{center}
  \vskip-0.3cm\caption{\small 
    Background-subtracted 
    distributions of all angles for the data 
    (points with error bars) and for the  
    $1^{++}$ fit projections (solid histograms).
  \label{fig:angles1P}
  }
\end{figure}

\begin{figure}[tbp]
  \begin{center}
  \ifthenelse{\boolean{pdflatex}}{
    \includegraphics*[width=\figsize]{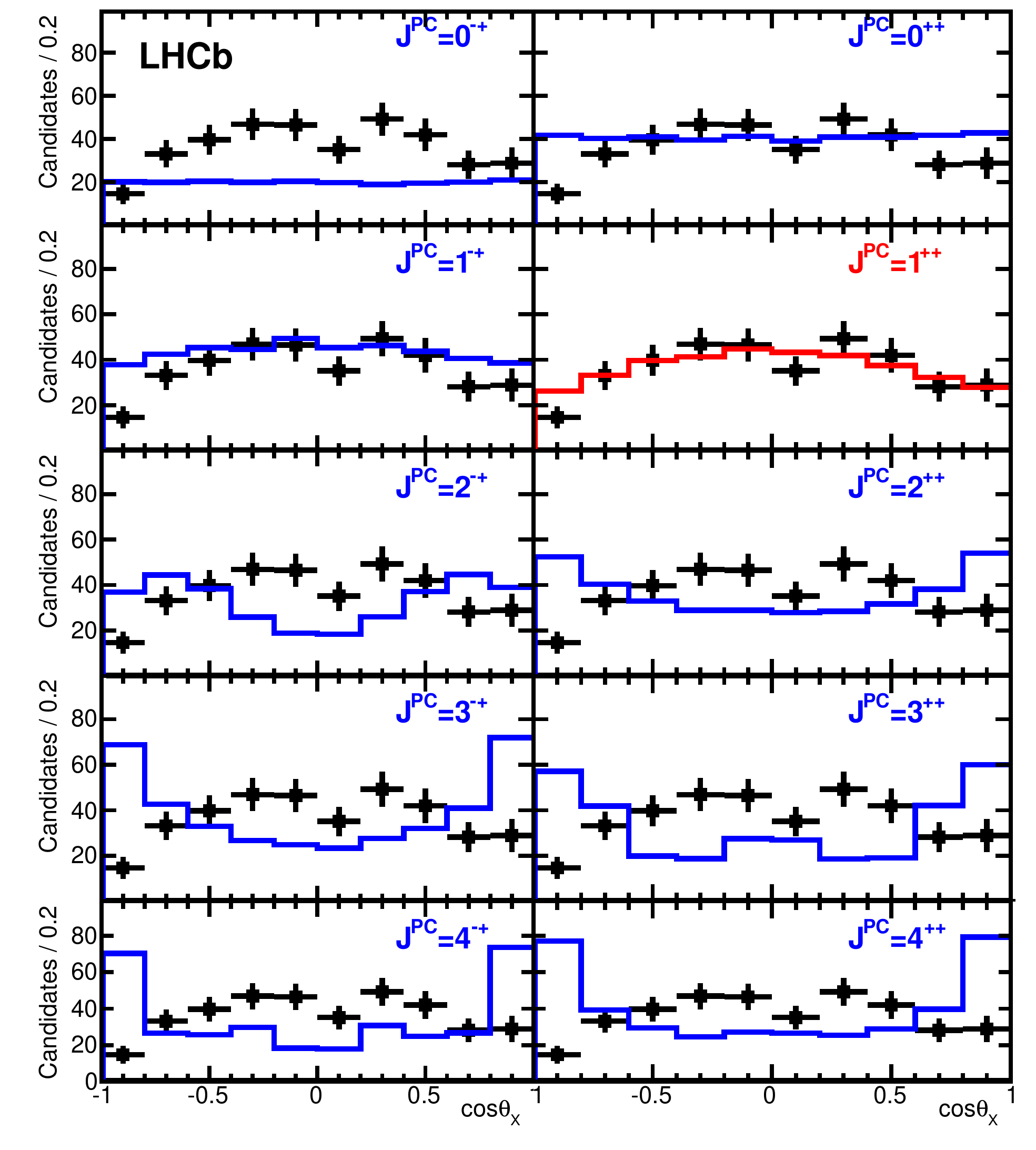}  
   }{
    \includegraphics*[width=\figsize]{histos_cosX_cosrho.eps}
   } 
  \end{center}
  \vskip-0.1cm\caption{\small
    Background-subtracted 
    distribution of $\cos\theta_X$ for 
    candidates with $|\cos\theta_{\rho}|>0.6$ 
    for the data (points with error bars)
    compared to the expected distributions 
    for various $\xff$ $J^{PC}$ assignments (solid histograms)
    with the $B_{LS}$ amplitudes obtained by the
    fit to the data in the five-dimensional angular space.
    The fit displays are normalized to the observed number of 
    the signal events in the full angular phase space. 
  \label{fig:cosThetaXwithThetaRhoCut}
  }
\end{figure}

The likelihood ratio 
$t \equiv -2\ln[\L(J_X^{\rm alt})/\L(1^{++})]$
is used as a test variable
to discriminate between the $1^{++}$ and 
alternative spin hypotheses considered ($J_X^{\rm alt}$).
The values of $t$ in the data ($t_{\rm data}$)
are positive, favoring the $1^{++}$ assignment.
They are incompatible with the distributions of $t$ observed in experiments 
simulated under various $J_X^{\rm alt}$ hypotheses, as
illustrated in Fig.~\ref{fig:spinana}. 
To quantify these disagreements we calculate the approximate significance of rejection
(p-value) of $J_X^{\rm alt}$ as %$n_\sigma(J_X^{\rm hyp})=
$(t_{\rm data}-\langle t \rangle)/\sigma(t)$, where
$\langle t \rangle$ and $\sigma(t)$ are the mean and rms deviations 
of the $t$ distribution under the $J_X^{\rm alt}$ hypothesis.
In all spin configurations tested, we exclude the alternative spin hypothesis 
with a significance of more than 16 standard deviations.
Values of $t$ in data are consistent with those expected in the $1^{++}$ case 
as shown in Fig.~\ref{fig:spinana}, 
with fractions of simulated $1^{++}$ experiments with $t< t_{\rm data}$ 
in the 25\%--91\%\ range.
Projections of the data and of the fit $\PDF$ onto individual angles show  
good consistency with the $1^{++}$ assignment 
as illustrated in Fig.~\ref{fig:angles1P}.
Inconsistency with the other assignments is apparent when 
correlations between various angles are exploited.
For example, the data projection onto $\cos\theta_X$ is consistent
only with the $1^{++}$ fit projection after requiring 
$|\cos\theta_{\rho}|>0.6$ (see Fig.~\ref{fig:cosThetaXwithThetaRhoCut}),
while inconsistency with the other quantum number assignments is less clear 
without the $\cos\theta_{\rho}$ requirement.   
  
The selection criteria are varied to probe for possible
biases from the background subtraction and the efficiency
corrections. 
By requiring $Q<0.1$ GeV, 
the background level is reduced by more than a factor of two, 
while losing only $20\%$ of the signal. 
By tightening the requirements on the $\pt$ of $\pi$, $K$ and $\mu$ candidates,
we decrease the signal efficiency 
by around 75\%\
with similar reduction
in the background level.
In all cases, the significance of the rejection of the disfavored
hypotheses is compatible with that expected from the simulation.
Likewise, the best fit $f_{\rm D}$ values determined for 
these subsamples of data change within the expected statistical fluctuations and
remain consistent with the upper limit we have set.

%\begin{table}[htbp]
%\caption{Results of the likelihood ratio test based on $-2\ln[\L(J_X^{\rm alt})/\L(1^{++})]$
%(see also Fig.~\ref{fig:spinana}).}
%\begin{center}
%\label{tab:spinana}
%\begin{tabular}{c|cc}
%\hline
% $J_{X}^{\rm alt}$ & $n_{\sigma}(J_{X}^{\rm alt})$ & $CL(1^{++})$ \\
%\hline
% $0^{-+}$ & $26\sigma$ & $91\%$  \\
% $0^{++}$ & $21\sigma$ & $60\%$  \\
% $1^{-+}$ & $22\sigma$ & $25\%$  \\
% $2^{-+}$ & $16\sigma$ & $65\%$  \\
% $2^{++}$ & $22\sigma$ & $84\%$  \\
% $3^{-+}$ & $23\sigma$ & $55\%$  \\
% $3^{++}$ & $21\sigma$ & $96\%$  \\
% $4^{-+}$ & $16\sigma$ & $28\%$  \\
% $4^{++}$ & $21\sigma$ & $61\%$  \\
%\hline
%\end{tabular}
%\end{center}
%\end{table}

In summary, 
the analysis of the angular correlations in $B^+\to\xff K^+$, $\xppj$, $\jmm$ decays,
performed for the first time without any assumption about 
the orbital angular momentum
in the $\xff$ decay,
confirms that the eigenvalues of total angular momentum, 
parity and charge-conjugation of the $X(3872)$ state
are $1^{++}$.
These quantum numbers are consistent with those predicted by the molecular or tetraquark models and 
with the $\chi_{c1}(2^3{\rm P}_1)$ charmonium state \cite{Achasov:2015oia}, possibly mixed 
with a molecule \cite{Hanhart:2011jz}. 
Other charmonium states are excluded.  
No significant D--wave fraction is found,
with an upper limit of  $4\%$ at 95\%~C.L.
The S--wave dominance is expected in the charmonium or tetraquark models,
in which the $\xff$ state has a compact size.
An extended size, as that predicted by the molecular model, 
implies more favorable conditions for the D wave. 
However, conclusive discrimination among models is difficult 
because quantitative predictions are not available.

\quad\newline
%\section*{Acknowledgements}
 
\noindent We express our gratitude to our colleagues in the CERN
accelerator departments for the excellent performance of the LHC. We
thank the technical and administrative staff at the LHCb
institutes. We acknowledge support from CERN and from the national
agencies: CAPES, CNPq, FAPERJ and FINEP (Brazil); NSFC (China);
CNRS/IN2P3 (France); BMBF, DFG, HGF and MPG (Germany); INFN (Italy); 
FOM and NWO (The Netherlands); MNiSW and NCN (Poland); MEN/IFA (Romania); 
MinES and FANO (Russia); MinECo (Spain); SNSF and SER (Switzerland); 
NASU (Ukraine); STFC (United Kingdom); NSF (USA).
The Tier1 computing centres are supported by IN2P3 (France), KIT and BMBF 
(Germany), INFN (Italy), NWO and SURF (The Netherlands), PIC (Spain), GridPP 
(United Kingdom).
We are indebted to the communities behind the multiple open 
source software packages on which we depend. We are also thankful for the 
computing resources and the access to software R\&D tools provided by Yandex LLC (Russia).
Individual groups or members have received support from 
EPLANET, Marie Sk\l{}odowska-Curie Actions and ERC (European Union), 
Conseil g\'{e}n\'{e}ral de Haute-Savoie, Labex ENIGMASS and OCEVU, 
R\'{e}gion Auvergne (France), RFBR (Russia), XuntaGal and GENCAT (Spain), Royal Society and Royal
Commission for the Exhibition of 1851 (United Kingdom).

\addcontentsline{toc}{section}{References}
\bibliographystyle{LHCb}
\bibliography{main,LHCb-PAPER,LHCb-CONF,LHCb-DP}

%\end{document} 

%\input{justification}

\newpage

% Author List ----------------------------
%  You need to get a new author list!
%%%%%%%%%%%%%%%%%%%%%%%%%%%%%%%%%%%%%%%%%%
\centerline{\large\bf LHCb collaboration}
\begin{flushleft}
\small
R.~Aaij$^{38}$, 
B.~Adeva$^{37}$, 
M.~Adinolfi$^{46}$, 
A.~Affolder$^{52}$, 
Z.~Ajaltouni$^{5}$, 
S.~Akar$^{6}$, 
J.~Albrecht$^{9}$, 
F.~Alessio$^{38}$, 
M.~Alexander$^{51}$, 
S.~Ali$^{41}$
% ...

{\footnotesize \it
%$\dagger$ deceased\\[1ex]
% ...
$ ^{5}$Clermont Universit\'{e}, Universit\'{e} Blaise Pascal, CNRS/IN2P3, LPC, Clermont-Ferrand, France\\
$ ^{6}$CPPM, Aix-Marseille Universit\'{e}, CNRS/IN2P3, Marseille, France\\
% ...
$ ^{9}$Fakult\"{a}t Physik, Technische Universit\"{a}t Dortmund, Dortmund, Germany\\
% ...
$ ^{37}$Universidad de Santiago de Compostela, Santiago de Compostela, Spain\\
$ ^{38}$European Organization for Nuclear Research (CERN), Geneva, Switzerland\\
% ...
$ ^{41}$Nikhef National Institute for Subatomic Physics, Amsterdam, The Netherlands\\
% ...
$ ^{46}$H.H. Wills Physics Laboratory, University of Bristol, Bristol, United Kingdom\\
% ...
$ ^{51}$School of Physics and Astronomy, University of Glasgow, Glasgow, United Kingdom\\
$ ^{52}$Oliver Lodge Laboratory, University of Liverpool, Liverpool, United Kingdom\\
% ...
}
\end{flushleft}
%%%%%%%%%%%%%%%%%%%%%%%%%%%%%%%%%%%%%%%%%%

%\input{supplemental}

\end{document}